\documentclass[a4paper,fleqn]{cas-dc}

\usepackage{amsmath}
\usepackage{amssymb}
\usepackage{bm}
\usepackage{color}
\usepackage{graphicx}
\usepackage{float}
\usepackage{listings}
\usepackage[numbers,sort&compress]{natbib}

\hypersetup{
  pdfpagemode=UseNone,
  pdfstartview=FitH,
  pdfstartpage=1,
  pdftitle={Sub-Second Collisionless Gyrokinetic Eigenvalue Solutions via Orbit-Invariant Decomposition},
  pdfauthor={Anrui Luo, Jingyi Yu, Huasheng Xie, and Jian Bao}
}

\graphicspath{{./}}

\newcommand{\iu}{\mathrm{i}}
\newcommand{\code}[1]{\texttt{#1}}

\begin{document}
\let\WriteBookmarks\relax
\def\floatpagepagefraction{1}
\def\textpagefraction{.001}

\shorttitle{Sub-Second Collisionless Gyrokinetic Eigenvalue Solutions}
\shortauthors{Luo et al.}

\title[mode=title]{Sub-Second Collisionless Gyrokinetic Eigenvalue Solutions via Orbit-Invariant Decomposition}

\author[1,2]{Anrui Luo}
\author[1,3]{Jingyi Yu}
\author[1]{Huasheng Xie}
\cormark[1]
\ead{huashengxie@gmail.com}
\ead{huashengxie@veloalpha.cn}
\author[1]{Jian Bao}

\affiliation[1]{organization={Beijing VeloAlpha Technology Co., Ltd.},
  city={Beijing},
  postcode={100080},
  country={China}}

\affiliation[2]{organization={School of Nuclear Science and Technology,
  University of Science and Technology of China},
  city={Hefei},
  postcode={230026},
  country={China}}

\affiliation[3]{organization={School of Physics, Peking University},
  city={Beijing},
  postcode={100871},
  country={China}}

\cortext[1]{Corresponding author}

\begin{abstract}
Fast analysis of microscopic drift-wave instabilities based on linear gyrokinetic simulations is desirable for modeling anomalous transport in fusion devices.
 In this work, we present an orbit-invariant decomposition method
for solving collisionless gyrokinetic eigenvalue problems. By discretizing velocity
space along orbit invariants using particle energy and magnetic moment,
the full eigenvalue matrix is separated into independent orbit blocks that
couple with each other through the field equation, greatly reducing both matrix dimension and
computational cost without sacrificing physics. Based on this method, we extend the MGK code
[Phys.\ Plasmas 24, 072106 (2017)] with both CPU and GPU implementations,
supporting collisionless electrostatic and electromagnetic linear simulations in
$s$--$\alpha$ and Miller equilibrium models.
For kinetic ion temperature gradient (ITG) and trapped electron mode (TEM)
eigenvalue problems, the solver reduces single-solution times to the
0.01--0.1~s range---more than three orders of magnitude faster
than CGYRO on the same hardware---enabling efficient large-scale parameter
scans. For fully electromagnetic KBM cases, it also achieves a speedup of
three orders of magnitude over CGYRO and HD7. The eigenfrequencies and mode structures are verified by comparing with CGYRO results. The method is generally applicable to all collisionless gyrokinetic
eigenvalue formulations and has been extended to fully electromagnetic
simulations.
\end{abstract}


\maketitle

\section{Introduction}
\label{sec:introduction}

Plasma pressure-gradient-driven microinstabilities can lead to anomalous transport, which
significantly degrades the confinement performance in fusion devices. Linear
gyrokinetic simulations~\cite{candy2003eulerian,candy2016cgyro,gorler2011gene,
peeters2009gkw} provide fundamental information such as mode structures, growth
rates, and frequencies~\cite{rewoldt2007linear}, and have been widely used for studying microinstabilities~\cite{brizard2007foundations,
krommes2012gyrokinetic,garbet2010gyrokinetic}. In addition, recent advances in
machine learning have made surrogate models possible for fusion research, and
training such models likewise requires a large high-fidelity data set~\cite{vandeplassche2020fast}. However, calculations with  well-established
gyrokinetic codes such as CGYRO~\cite{candy2016cgyro},
GENE~\cite{gorler2011gene}, and
HD7~\cite{dong1987finitebeta,wang2012linear} haven't been optimized for greatly reducing the computational time~\cite{xie2017comparisons,roman2010fast}.

Many valuable studies have achieved excellent results in accelerating linear
simulations, including TGLF\cite{staebler2007tglf} and QuaLiKiz~\cite{bourdelle2016qualikiz}, as well as the dispersion model of Morren and related
reduced kinetic formulations~\cite{ivanov2023dispersion,frei2023moment,
morren2026fast}.
However, these approaches generally employ reduced-order models and do not
independently solve the complete gyrokinetic problem with a self-consistent
kinetic--field closure. For eigenvalue problems, Schur
elimination and matrix-free operators have substantially accelerated
computation~\cite{xie2017comparisons,roman2010fast}. Such matrix problems also
often achieve better performance on GPUs~\cite{abdelfattah2021batched}.

For the linear collisionless gyrokinetic model considered here,
each independent guiding center evolves along its corresponding unperturbed orbit, so various orbit-type contributions to particle response are connected only through the field equations. Regarding the
eigenvalue problem, we discretize the perturbed distribution in velocity space using
particle energy and magnetic moment, which can effectively separate the
kinetic matrix into independent matrix blocks, for each
velocity-space quadrature point. The dimension of each orbit matrix is therefore
smaller than that of the complete matrix with vpara and vperp coordinates by a factor of
$N_{\mathrm{orb}}$, and retains an algebraically equivalent
solution of the coupled problem. Using this method, we extended and accelerated
the MGK code~\cite{xie2017comparisons}.

The remainder of this work is organized as follows.
Section~\ref{sec:model} introduces the basic gyrokinetic theory and the
$s$--$\alpha$ and Miller geometries used by the code.
Section~\ref{sec:numerical_formulation} describes the numerical discretization.
Section~\ref{sec:results} reproduces several typical simulations with MGK and
compares the results with the established code CGYRO.
Section~\ref{sec:performance} reports the measured computational performance.
Appendix~\ref{app:electromagnetic} describes the electromagnetic model and
reports its validation and computational performance.
Finally, Sec.~\ref{sec:conclusions} summarizes the work and discusses future
directions.

\section{Physical model}
\label{sec:model}

\subsection{Linear gyrokinetic model}

Magnetically confined fusion plasmas contain multiple particle species, with
number densities reaching $10^{20}\,\mathrm{m}^{-3}$. Resolving their collective
behavior generally requires a kinetic description. For each species $s$, the
distribution function $f_s(\mathbf{x},\mathbf{v},t)$ is defined on the
six-dimensional phase space of particle position $\mathbf{x}$ and velocity
$\mathbf{v}$. It satisfies
\begin{equation}
  \frac{\partial f_s}{\partial t}
  + \mathbf{v}\cdot\nabla_{\mathbf{x}} f_s
  + \frac{q_s}{m_s}
    (\mathbf{E}+\mathbf{v}\times\mathbf{B})\cdot\nabla_{\mathbf{v}} f_s
  = C[f_s],
  \label{eq:kinetic_equation}
\end{equation}
where $q_s$ and $m_s$ are the charge and mass, $\mathbf{E}$ and $\mathbf{B}$ are
the electric and magnetic fields, and $C[f_s]$ is the collision operator.

Particles in tokamak plasmas are strongly magnetized and undergo rapid
gyromotion about the magnetic field. The gyrokinetic ordering assumes
~\cite{brizard2007foundations,krommes2012gyrokinetic}
\begin{equation}
  \epsilon_{\mathrm{GK}}
  \sim \frac{\rho_{\mathrm{th},s}}{L}
  \sim \frac{\omega}{\Omega_s}
  \sim \frac{k_\parallel}{k_\perp}
  \ll 1,
  \label{eq:gyrokinetic_ordering}
\end{equation}
where $\rho_{\mathrm{th},s}$ is the gyroradius, $L$ is an equilibrium scale length, $\omega$
is a characteristic fluctuation frequency, $\Omega_s$ is the gyrofrequency,
and $k_\parallel$ and $k_\perp$ are the parallel and perpendicular
wavenumbers. Averaging over the fast gyrophase then reduces the distribution
function to $F_s(\mathbf{R},E_s,\mu,\sigma,t)$, where
$\mathbf{R}$ is the gyrocenter position, $E_s$ is the gyrocenter
energy, $\mu$ is the magnetic moment, and $\sigma=\operatorname{sgn}(v_\parallel)$
labels the direction of parallel motion.

Although collisions can modify microinstability growth rates and mode
structure, this work adopts the collisionless limit. The gyrocenter
distribution therefore satisfies
\begin{equation}
  \frac{\partial F_s}{\partial t}
  + \{F_s,H_s\}_{\mathrm{gc}} = 0,
  \label{eq:gyrokinetic_equation}
\end{equation}
where $H_s$ is the gyrocenter Hamiltonian and
$\{\cdot,\cdot\}_{\mathrm{gc}}$ is the gyrocenter Poisson bracket. We decompose
the Hamiltonian and distribution function into equilibrium and perturbed parts,
\begin{equation}
  H_s = H_{0s} + \delta H_s,
  \qquad
  F_s = F_{0s} + \delta F_s.
  \label{eq:hamiltonian_distribution_decomposition}
\end{equation}
and retain terms that are first order in the perturbations. For any phase-space
function $G$, define the derivative along an unperturbed gyrocenter orbit by
\begin{equation}
  \frac{D_{0s}G}{Dt}
  \equiv \frac{\partial G}{\partial t}
  + \{G,H_{0s}\}_{\mathrm{gc}}.
  \label{eq:unperturbed_orbit_derivative}
\end{equation}
The linearized gyrokinetic equation is then
\begin{equation}
  \frac{D_{0s}\delta F_s}{Dt}
  = -\{F_{0s},\delta H_s\}_{\mathrm{gc}}.
  \label{eq:linearized_gk_deltaf_initial}
\end{equation}

The equilibrium distribution is Maxwellian,
\begin{equation}
  F_{0s}(\mathbf{R},E_s)
  =\frac{n_s(\mathbf{R})}{(2\pi T_s(\mathbf{R})/m_s)^{3/2}}
   \exp\!\left(-\frac{E_s}{T_s(\mathbf{R})}\right),
  \label{eq:equilibrium_maxwellian}
\end{equation}
For the unperturbed Hamiltonian $H_{0s}=E_s$, the chain rule gives
\begin{equation}
  \begin{split}
  \{F_{0s},\delta H_s\}_{\mathrm{gc}}
  ={}&
  \left.\frac{\partial F_{0s}}{\partial H_{0s}}
  \right|_{\mathbf{R},\mu}
  \{H_{0s},\delta H_s\}_{\mathrm{gc}} \\
  &+
  \left.\nabla_{\mathbf{R}}F_{0s}\right|_{H_{0s},\mu}
  \cdot\{\mathbf{R},\delta H_s\}_{\mathrm{gc}}.
  \end{split}
  \label{eq:maxwellian_bracket_expansion}
\end{equation}
The energy derivative of the Maxwellian is
$\partial F_{0s}/\partial H_{0s}=-F_{0s}/T_s$.
The second term contains the perturbed gyrocenter velocity
$\delta\dot{\mathbf{R}}_s\equiv
\{\mathbf{R},\delta H_s\}_{\mathrm{gc}}$. For the electrostatic model considered
here, $\delta H_s=q_s\langle\phi\rangle_{\mathbf R}$, where $\langle\phi\rangle_{\mathbf R}$ is the gyroaveraged
electrostatic potential. Its perpendicular contribution is the
$\mathbf{E}\times\mathbf{B}$ drift
\begin{equation}
  \mathbf{v}_{E}
  \equiv \delta\dot{\mathbf{R}}_{s\perp}
  = \frac{\hat{\mathbf b}\times\nabla\langle\phi\rangle_{\mathbf R}}{B},
  \qquad \hat{\mathbf b}=\frac{\mathbf{B}}{B}.
  \label{eq:vchi_definition}
\end{equation}
Consequently,
\begin{equation}
  \{F_{0s},\delta H_s\}_{\mathrm{gc}}
  = \frac{F_{0s}}{T_s}
    \{\delta H_s,H_{0s}\}_{\mathrm{gc}}
  + \left.\mathbf{v}_{E}\cdot
    \nabla_{\mathbf{R}}F_{0s}\right|_{H_{0s},\mu},
  \label{eq:maxwellian_bracket_final}
\end{equation}
and Eq.~\eqref{eq:linearized_gk_deltaf_initial} becomes
\begin{equation}
  \frac{D_{0s}\delta F_s}{Dt}
  = -\frac{F_{0s}}{T_s}
    \{\delta H_s,H_{0s}\}_{\mathrm{gc}}
  - \left.\mathbf{v}_{E}\cdot
    \nabla_{\mathbf{R}}F_{0s}\right|_{H_{0s},\mu}.
  \label{eq:linearized_gk_deltaf}
\end{equation}

We now define the nonadiabatic gyrocenter response by
\begin{equation}
  h_s \equiv \delta F_s
  + \frac{F_{0s}}{T_s}\delta H_s.
  \label{eq:hs_definition}
\end{equation}
Using Eq.~\eqref{eq:linearized_gk_deltaf} and the stationarity of the
equilibrium yields
\begin{equation}
  \frac{D_{0s}h_s}{Dt}
  = \frac{F_{0s}}{T_s}
    \frac{\partial\delta H_s}{\partial t}
  - \left.\mathbf{v}_{E}\cdot
    \nabla_{\mathbf{R}}F_{0s}\right|_{H_{0s},\mu}.
  \label{eq:linearized_gk_h}
\end{equation}

It remains to expand the unperturbed orbit derivative in the phase-space
coordinates used by the solver. Because $H_{0s}=E_s$ in a static
equilibrium, both $E_s$ and $\mu$ are constants along each
unperturbed orbit branch. Therefore,
\begin{equation}
  \frac{D_{0s}}{Dt}
  = \frac{\partial}{\partial t}
  + \left.(v_\parallel\hat{\mathbf b}+\mathbf{v}_{ds})\cdot
    \nabla_{\mathbf{R}}\right|_{E_s,\mu,\sigma},
  \label{eq:orbit_derivative_energy_coordinates}
\end{equation}
where
\begin{equation}
  \begin{aligned}
    v_\parallel
    &= \sigma\sqrt{\frac{2}{m_s}(E_s-\mu B)}, \\
    \mathbf{v}_{ds}
    &= \frac{\hat{\mathbf b}}{q_sB}\times
      \left(m_sv_\parallel^2\boldsymbol{\kappa}+\mu\nabla B\right), \\
    \boldsymbol{\kappa}
    &= \hat{\mathbf b}\cdot\nabla\hat{\mathbf b}.
  \end{aligned}
  \label{eq:parallel_velocity_and_magnetic_drift}
\end{equation}
At a bounce point, $\sigma$ changes sign through the orbit boundary condition;
there is no separate mirror-force derivative with respect to $v_\parallel$ in
the energy-coordinate representation. The resulting equation is
\begin{equation}
  \begin{aligned}
    &\left[
      \frac{\partial}{\partial t}
      + (v_\parallel\hat{\mathbf b}+\mathbf{v}_{ds})\cdot
        \left.\nabla_{\mathbf{R}}\right|_{E_s,\mu,\sigma}
    \right]h_s \\
    &\qquad= \frac{F_{0s}}{T_s}
      \frac{\partial\delta H_s}{\partial t} \\
    &\qquad\quad- \left.\mathbf{v}_{E}\cdot
      \nabla_{\mathbf{R}}F_{0s}\right|_{E_s,\mu}.
  \end{aligned}
  \label{eq:linearized_gk_energy_coordinates}
\end{equation}

\subsection{Gyroaveraging and quasineutrality closure}
\label{sec:gyroaveraging_and_quasineutrality}

For a local perpendicular Fourier component, $\mathbf{k}_\perp$ is taken as
constant across one gyro-ring. Averaging over the gyrophase $\vartheta$ at fixed
$\mathbf R$ gives
\begin{equation}
  \begin{aligned}
    \langle\phi\rangle_{\mathbf R}(\mathbf R,t)
    &=\frac{1}{2\pi}\int_0^{2\pi}
      \phi\!\left(\mathbf R+\boldsymbol\rho_s(\vartheta),t\right)
      \,d\vartheta \\
    &=\phi(\mathbf R,t)\frac{1}{2\pi}\int_0^{2\pi}
      \exp\!\left[\mathrm i\mathbf k_\perp\cdot
      \boldsymbol\rho_s(\vartheta)\right]d\vartheta \\
    &=J_0(a_s)\phi(\mathbf R,t).
  \end{aligned}
  \label{eq:gyroaveraged_potential}
\end{equation}
where $a_s=k_\perp v_\perp/|\Omega_s|$.

The field equation involves the charge density at fixed particle position,
whereas $h_s$ is defined at fixed gyrocenter position. To first order, the
pullback to particle phase space gives
$\delta f_s=\langle h_s\rangle_{\mathbf{x}}
-q_sF_{0s}\phi/T_s$. Taking its velocity-space moment and applying
the inverse gyroaverage gives
\begin{equation}
  \delta n_s
  =\int d^3v\,J_0(a_s)h_s
  -\frac{q_sn_s}{T_s}\phi.
  \label{eq:particle_density_response}
\end{equation}

Let $N_{\mathrm{sp}}$ be the total number of species and $N_{\mathrm{kin}}$ the
number treated kinetically, with kinetic species indexed first. For an
adiabatic species, $h_s=0$, while its Boltzmann response remains in the density. In the
limit $k_\perp^2\lambda_D^2\ll1$, substituting
Eq.~\eqref{eq:particle_density_response} into Poisson's equation,
$-\epsilon_0\nabla^2\phi=\sum_s q_s\delta n_s$, gives the quasineutrality
condition
\begin{equation}
  \sum_{s=1}^{N_{\mathrm{kin}}}q_s
  \int d^3v\,J_0(a_s)h_s
  -\sum_{s=1}^{N_{\mathrm{sp}}}
  \frac{q_s^2n_s}{T_s}\phi
  =0.
  \label{eq:continuous_quasineutrality}
\end{equation}
Together with Eq.~\eqref{eq:linearized_gk_energy_coordinates}, this condition
closes the continuous electrostatic model for $h_s$ and $\phi$.

\subsection{Local magnetic geometry}
\label{sec:local_magnetic_geometry}

Evaluation of Eqs.~\eqref{eq:linearized_gk_energy_coordinates}--%
\eqref{eq:continuous_quasineutrality} requires the field strength, parallel
metric, magnetic-drift projection, and perpendicular wavenumber along the
field line. These coefficients are obtained from either the analytic
$s$--$\alpha$ model or a local Miller equilibrium.

For the circular $s$--$\alpha$ model, the geometric coefficients are taken
as~\cite{connor1978shear}
\begin{equation}
  \begin{gathered}
    \widehat B(\theta)\equiv\frac{B(\theta)}{B_0}
      =\frac{1}{1+\epsilon\cos\theta},
    \qquad
    \hat{\mathbf b}\cdot\nabla
      =\frac{1}{qR_0}\frac{\partial}{\partial\theta},\\
    \frac{k_x(\theta)}{k_y}
      =\hat{s}(\theta-\theta_0)-\alpha\sin\theta,\\
    k_\perp^2(\theta)=k_y^2\left[1+\frac{k_x^2(\theta)}{k_y^2}\right],
    \qquad
    \mathbf{k}_\perp\cdot\mathbf{v}_{ds}
      \propto\cos\theta+\frac{k_x}{k_y}\sin\theta.
  \end{gathered}
  \label{eq:s_alpha_geometry}
\end{equation}
Here $r$ is the local minor-radius label, $R_0$ and $B_0$ are the reference
major radius and magnetic-field strength, $\epsilon=r/R_0$, $q$ is the safety
factor, and $\hat{s}=(r/q)dq/dr$ is the magnetic shear. The pressure-gradient
parameter is $\alpha=-(2\mu_0q^2R_0/B_0^2)dp/dr$; $\theta_0$ is the ballooning angle and
$k_y$ is the prescribed binormal wavenumber. The final relation in
Eq.~\eqref{eq:s_alpha_geometry} displays the geometric factor entering
$\mathbf{k}_\perp\cdot\mathbf{v}_{ds}$, while
$k_x/k_y$ describes the radial component of the sheared perpendicular
wavevector.

For a shaped flux surface, the Miller model uses the local parameterization
\begin{equation}
  \begin{aligned}
    R(r,\theta)
      &=R_0(r)+r\cos\!\left[
        \theta+\arcsin[\delta(r)]\sin\theta\right],\\
    Z(r,\theta)&=\kappa(r)r\sin\theta,\\
    \Delta'&=\frac{dR_0}{dr},
    \qquad
    s_\kappa=\frac{r}{\kappa}\frac{d\kappa}{dr},
    \qquad
    s_\delta=r\frac{d\delta}{dr}.
  \end{aligned}
  \label{eq:miller_surface}
\end{equation}
Here $\kappa$ and $\delta$ are the elongation and triangularity, $\Delta'$ is
the radial derivative of the Shafranov shift, and $s_\kappa$ and $s_\delta$
describe the radial variation of the shape; the definition of $s_\delta$
follows the GACODE convention~\cite{candy2016cgyro}. Together with $q$, $\hat{s}$,
$\alpha$, and the prescribed $(k_y,\theta_0)$, these quantities determine
$B(\theta)$, the parallel metric, $\nabla B$, $\boldsymbol\kappa$, and
$k_\perp(\theta)$ through the local equilibrium
construction~\cite{miller1998local}.

For either geometry, the selected field-strength profile determines
$v_\parallel$ and the turning points in
Eq.~\eqref{eq:parallel_velocity_and_magnetic_drift}; the parallel metric,
$\nabla B$, and $\boldsymbol\kappa$ determine the streaming and magnetic-drift
terms in Eq.~\eqref{eq:linearized_gk_energy_coordinates}. Finally,
$k_\perp(\theta)$ enters the gyroaveraging argument in
Eq.~\eqref{eq:gyroaveraged_potential}, with the local gyroradius
$v_\perp/|\Omega_s(\theta)|$ carrying the corresponding
$B(\theta)$ dependence, and therefore enters the density response and
quasineutrality closure in
Eqs.~\eqref{eq:particle_density_response} and
\eqref{eq:continuous_quasineutrality}. Both geometries therefore provide
alternative spatial coefficient sets for the same closed kinetic--field
system.

\section{Numerical formulation}
\label{sec:numerical_formulation}

\subsection{Complementary distribution and continuous normal modes}
\label{sec:continuous_normal_modes}

The time-dependent terms in
Eq.~\eqref{eq:linearized_gk_energy_coordinates} combine as
\begin{equation}
  \frac{\partial h_s}{\partial t}
  -\frac{F_{0s}}{T_s}
    \frac{\partial\delta H_s}{\partial t}
  =\frac{\partial}{\partial t}
    \left(h_s-\frac{F_{0s}}{T_s}\delta H_s\right).
  \label{eq:hs_time_derivative_combination}
\end{equation}
Naturally, we have
\begin{equation}
  g_s
  \equiv h_s-\frac{F_{0s}}{T_s}\delta H_s
  =h_s-\frac{q_sF_{0s}}{T_s}\langle\phi\rangle_{\mathbf R}
  =\delta F_s.
  \label{eq:complementary_distribution_dimensional}
\end{equation}
which collects all explicit time dependence. The linearized local system is
invariant under time translations, so normal-mode solutions can be sought in
the form
\begin{equation}
  \begin{aligned}
    g_s&=\widehat g_s(\theta,E_s,\mu,\sigma)
      \exp[\mathrm{i}\mathbf k_\perp\cdot\mathbf R-\mathrm{i}\omega t],\\
    \phi&=\widehat\phi(\theta)
      \exp[\mathrm{i}\mathbf k_\perp\cdot\mathbf R-\mathrm{i}\omega t].
  \end{aligned}
  \label{eq:normal_mode_ansatz}
\end{equation}
Here $\mathbf k_\perp$ is the prescribed local perpendicular wavevector,
$\widehat g_s$ and $\widehat\phi$ are the mode amplitudes, and
$\omega=\omega_r+\mathrm{i}\gamma$, with $\omega_r$ the oscillation frequency
and $\gamma$ the growth rate. The eigenproblem generally admits several normal
modes. The mode with the largest $\gamma$ grows most rapidly and typically
dominates the linear evolution of a tokamak microinstability.

Substituting Eq.~\eqref{eq:complementary_distribution_dimensional} into
Eq.~\eqref{eq:linearized_gk_energy_coordinates}, and using
Eq.~\eqref{eq:gyroaveraged_potential}, gives, with spatial derivatives taken at
fixed $(E_s,\mu,\sigma)$,
\begin{equation}
  \begin{aligned}
    &\left[-\mathrm{i}\omega
      +v_\parallel\nabla_\parallel
      +\mathrm{i}\mathbf k_\perp\cdot\mathbf v_{ds}\right]\\
    &\qquad\times
      \left[g_s+\frac{q_sF_{0s}}{T_s}J_0(a_s)\phi\right]\\
    &\qquad=-\mathrm{i}\frac{q_sF_{0s}}{T_s}
      (\omega-\omega_{\ast s}^{T})J_0(a_s)\phi.
  \end{aligned}
  \label{eq:dimensional_g_equation}
\end{equation}
The equilibrium-gradient frequency in Eq.~\eqref{eq:dimensional_g_equation} is
\begin{equation*}
  \begin{aligned}
    \omega_{\ast s}^{T}
    &=\frac{T_s}{q_sB}\mathbf k_\perp\cdot
      \left(\hat{\mathbf b}\times
      \left.\nabla_{\mathbf R}\ln F_{0s}
      \right|_{E_s,\mu}\right)\\
    \left.\nabla_{\mathbf R}\ln F_{0s}
      \right|_{E_s,\mu}
    &=
      \nabla_{\mathbf R}\ln n_s
      +\left(\frac{E_s}{T_s}-\frac{3}{2}\right)
       \nabla_{\mathbf R}\ln T_s.
  \end{aligned}
\end{equation*}

The field equation in the same unknown follows by substituting
$h_s=g_s+q_sF_{0s}J_0(a_s)\phi/T_s$ into
Eq.~\eqref{eq:continuous_quasineutrality}:
\begin{align}
  &\sum_{s=1}^{N_{\mathrm{kin}}}q_s\int d^3v\,J_0(a_s)g_s \nonumber\\
  &\quad+
  \left[
    \sum_{s=1}^{N_{\mathrm{kin}}}\frac{q_s^2}{T_s}
      \int d^3v\,F_{0s}J_0^2(a_s)
    -\sum_{s=1}^{N_{\mathrm{sp}}}\frac{q_s^2n_s}{T_s}
  \right]\phi=0.
  \label{eq:dimensional_g_quasineutrality}
\end{align}

\subsection{Normalized generalized eigenvalue problem}
\label{sec:normalized_eigenproblem}

The continuous eigenproblem is now expressed in the normalized variables used
by the solver. Write $q_s=Z_se$ and let $v_{\mathrm{th},s}=(T_s/m_s)^{1/2}$ for each
species. Let $R_0$ and $B_0$ be the reference major radius and magnetic-field
strength. With the main ion as reference, define $v_{\mathrm{th},i}=(T_i/m_i)^{1/2}$ and
$\rho_i=v_{\mathrm{th},i}m_i/(|Z_i|eB_0)$. For comparisons using the sound-speed
convention, define $c_s=(T_e/m_i)^{1/2}$ and
$\rho_s=c_sm_i/(|Z_i|eB_0)$. Frequencies are normalized by $v_{\mathrm{th},i}/R_0$,
perpendicular wavenumbers by $\rho_i^{-1}$, the potential by $T_i/(Z_i e)$,
and the velocity-space variables of each species by $v_{\mathrm{th},s}$. Both $F_{0s}$ and
$g_s$ are divided by $n_s/v_{\mathrm{th},s}^3$, so that the normalized Maxwellian satisfies
$\int d^3v\,F_{0s}=1$. The normalized species parameters are
\begin{equation}
  z_s=\frac{Z_s}{Z_i},
  \qquad
  \tau_s=\frac{T_s}{T_i},
  \qquad
  \bar n_s=\frac{n_s}{n_i}.
  \label{eq:normalized_species_parameters}
\end{equation}
These ratios enter explicitly in the kinetic and quasineutrality equations
below. The normalized distribution functions and mode
amplitudes retain their symbols; from this point onward, $F_{0s}$ denotes the
normalized Maxwellian, and
$\widehat\omega_{\ast s}^{T}=(R_0/v_{\mathrm{th},i})\omega_{\ast s}^{T}$.

Applying these normalizations to Eq.~\eqref{eq:dimensional_g_equation}
cancels the common field-response term proportional to
$\omega(z_s/\tau_s)F_{0s}J_0(a_s)\phi$. The dimensional orbit derivative then
appears in the combination $R_0/v_{\mathrm{th},i}$:
\begin{equation}
  \begin{aligned}
    &-\mathrm{i}\frac{R_0}{v_{\mathrm{th},i}}
      \left[v_\parallel\nabla_\parallel
      +\mathrm{i}\mathbf k_\perp\cdot\mathbf v_{ds}\right]\\
    &\qquad\times
      \left[g_s+\frac{z_s}{\tau_s}F_{0s}J_0(a_s)\phi\right]\\
    &\qquad-\frac{z_s}{\tau_s}F_{0s}\widehat\omega_{\ast s}^{T}
      J_0(a_s)\phi=\omega g_s.
  \end{aligned}
  \label{eq:implemented_gk}
\end{equation}
Normalizing Eq.~\eqref{eq:dimensional_g_quasineutrality} gives
\begin{align}
  &\sum_{s=1}^{N_{\mathrm{kin}}}z_s\bar n_s
    \int d^3v\,J_0(a_s)g_s \nonumber\\
  &\quad+
  \left[-\sum_{s=1}^{N_{\mathrm{sp}}}\frac{\bar n_sz_s^2}{\tau_s}
    +\sum_{s=1}^{N_{\mathrm{kin}}}\frac{\bar n_sz_s^2}{\tau_s}
      \int d^3v\,F_{0s}J_0^2(a_s)\right]\phi=0.
  \label{eq:implemented_quasineutrality}
\end{align}
The $F_{0s}J_0^2$ term is the gyroaveraged equilibrium response transferred to
the field closure by the definition of $g_s$.

The continuous equations can be written in terms of four linear operators. The
orbit term is represented by $\widehat L_s$, the field-driven kinetic term by
$\widehat D_s$, the gyroaveraged charge-density moment by $\widehat Q_s$, and
the field response by $\widehat P$, with the sign of $\widehat P$ inherited
from Eq.~\eqref{eq:implemented_quasineutrality}. The resulting equations are
\begin{equation}
  \widehat L_sg_s+\widehat D_s\phi=\omega g_s,
  \qquad
  \sum_{s=1}^{N_{\mathrm{kin}}}\widehat Q_sg_s+\widehat P\phi=0.
  \label{eq:discrete_kinetic_field_blocks}
\end{equation}
Collecting the unknowns
in $x=(g_1,\ldots,g_{N_{\mathrm{kin}}},\phi)^T$ gives
\begin{equation}
  \widehat Ax=\omega\widehat Bx,
  \qquad
  \begin{aligned}
    \widehat A&=
    \begin{bmatrix}
      \widehat L_1 &        &        & \widehat D_1\\
          & \ddots &        & \vdots\\
          &        & \widehat L_{N_{\mathrm{kin}}}& \widehat D_{N_{\mathrm{kin}}}\\
      \widehat Q_1 & \cdots & \widehat Q_{N_{\mathrm{kin}}}& \widehat P
    \end{bmatrix},\\
    \widehat B&=\operatorname{diag}
      (I_{\mathrm{kinetic}},0_{\mathrm{field}}).
  \end{aligned}
  \label{eq:generalized_eigenproblem}
\end{equation}
The complete kinetic time derivative produces the identity-operator block of
$\widehat B$, and the algebraic quasineutrality constraint produces the zero
field block. Thus $\widehat B$ is singular. The continuous generalized
eigenvalues $\omega$ are those for which $\widehat A-\omega\widehat B$ has a
nontrivial kernel; the corresponding vector $x$ contains the kinetic
distribution functions and electrostatic potential.

\subsection{Orbit discretization and Schur reduction}
\label{sec:phase_space_discretization}

The kinetic block $\mathsf L_s$ introduced in
Eq.~\eqref{eq:discrete_kinetic_field_blocks} is the discrete representation of
the continuous orbit operator
\begin{equation}
  \widehat L_s=
  -\mathrm{i}\frac{R_0}{v_{\mathrm{th},i}}
  \left[
    v_\parallel\hat{\mathbf b}\cdot\nabla\theta
      \left.\frac{\partial}{\partial\theta}
      \right|_{E_s,\mu,\sigma}
    +\mathrm{i}\mathbf k_\perp\cdot\mathbf v_{ds}
  \right].
  \label{eq:continuous_orbit_operator}
\end{equation}
A hat denotes a continuous orbit operator, whereas a sans-serif $\mathsf L$
denotes its discrete matrix representation.
This is the invariant-coordinate form established in
Eq.~\eqref{eq:orbit_derivative_energy_coordinates}. Its numerical role follows
by writing the same parallel Hamiltonian motion in
$(\theta,v_\parallel,\mu)$ coordinates:
\begin{equation}
  \begin{aligned}
  &v_\parallel\hat{\mathbf b}\cdot\nabla\theta
    \left.\frac{\partial}{\partial\theta}
    \right|_{v_\parallel,\mu}
  -\frac{\mu}{m_s}\hat{\mathbf b}\cdot\nabla B
    \frac{\partial}{\partial v_\parallel}
  \\
  &\qquad=
  v_\parallel\hat{\mathbf b}\cdot\nabla\theta
    \left.\frac{\partial}{\partial\theta}
    \right|_{E_s,\mu,\sigma}.
  \end{aligned}
  \label{eq:mirror_force_coordinate_identity}
\end{equation}
On a tensor-product $(\theta,v_\parallel)$ grid, the mirror-force derivative
in Eq.~\eqref{eq:mirror_force_coordinate_identity} connects different velocity
nodes. In the energy-coordinate form, it is incorporated into differentiation
along a curve of constant
$E_s=m_sv_\parallel^2/2+\mu B$. The variables $(E_s,\mu)$
therefore parameterize the curves on which $\widehat L_s$ acts.

For one accessible orbit $o$, let $\tau_o$ denote its trajectory
coordinate. A passing orbit has
$o=(E_s,\mu,\sigma)$ and $\tau_o=\theta$, whereas a trapped orbit
is labeled by $(E_s,\mu)$ and its magnetic well and includes both
signs of $v_\parallel$. Restricting Eq.~\eqref{eq:implemented_gk} to one such
orbit gives the continuous response
\begin{equation}
  \begin{aligned}
    (\widehat L_{so}-\omega)g_{so}
    ={}&-\widehat L_{so}
      \left(\frac{z_s}{\tau_s}F_{0s}J_0\phi\right)\\
     &+\frac{z_s}{\tau_s}F_{0s}\widehat\omega_{\ast s}^{T}J_0\phi .
  \end{aligned}
  \label{eq:continuous_orbit_response}
\end{equation}
No derivative with respect to $E_s$ or $\mu$ occurs in this equation.
The collisionless kinetic response is consequently a family of independent
one-dimensional equations, parameterized by the velocity-space invariants and
coupled through the common electrostatic field.

Let $B(\theta)$ denote the field used to classify the orbits,
with extrema $B_{\min}$ and $B_{\max}$. The invariant
pair is reparameterized for quadrature by the speed and
\begin{equation*}
  \lambda=\frac{\mu B_{\min}}{E_s}
  =1-\xi_0^2,
\end{equation*}
where $\xi_0$ is the magnitude of the pitch cosine at
$B_{\min}$. Quadrature rules constructed by the Jacobi-matrix
procedure of Golub and Welsch \cite{golub1969quadrature} are divided at the
separatrix
$\lambda B_{\max}/B_{\min}=1$. Passing pitch points are placed
directly in $\xi_0$, while trapped points are generated by mapping a
turning-point-angle rule to $\xi_0$; the corresponding pitch-map Jacobian is
included in the quadrature weights. This turning-point angle is distinct from
the periodic trajectory coordinate $\tau_b$ introduced below.

Let $j$ denote one resulting speed--pitch point. Each passing point produces
the two orbit blocks $o=(j,\sigma)$, while each trapped point produces one
closed block in every accessible magnetic well. The quadrature samples the
invariant-parameterized family in Eq.~\eqref{eq:continuous_orbit_response}.
When a response is deposited at a local field position, its reference weight
is multiplied by
\begin{equation*}
  J_{\mathrm{loc}}(\theta)=
  \frac{B(\theta)}{B_{\min}}
  \frac{|v_{\parallel0}|}{|v_\parallel(\theta)|},
\end{equation*}
which converts the invariant quadrature to the local velocity-space measure.

Discretizing the trajectory coordinate $\tau_o$ at $n_{so}$ nodes with the
orbit differentiation matrix $\mathsf T_o$ gives
\begin{equation}
  \mathsf L_{so}=
  -\mathrm{i}\frac{R_0}{v_{\mathrm{th},i}}
    \operatorname{diag}(\dot\tau_{so})\mathsf T_o
  +\frac{R_0}{v_{\mathrm{th},i}}
    \operatorname{diag}(\mathbf k_\perp\cdot\mathbf v_{dso}),
  \label{eq:discrete_orbit_operator}
\end{equation}
where the coefficients are evaluated at the nodes of that orbit. Since the
velocity quadrature labels are parameters in Eq.~\eqref{eq:continuous_orbit_response},
the discrete species block is explicitly block diagonal. Bold symbols denote
discrete coefficient vectors; the first such vector is
\begin{equation}
  \begin{aligned}
    \mathbf g_s&=
      \begin{bmatrix}
        \mathbf g_{s1}&\mathbf g_{s2}&\cdots&\mathbf g_{s,N_{\mathrm{orb},s}}
      \end{bmatrix}^{T},\\
    \mathsf L_s&=\operatorname{diag}\left(
      \mathsf L_{s1},\mathsf L_{s2},\ldots,\mathsf L_{s,N_{\mathrm{orb},s}}
    \right).
  \end{aligned}
  \label{eq:kinetic_orbit_direct_sum}
\end{equation}
where $N_{\mathrm{orb},s}$ is the number of orbit blocks for species $s$.
Thus $\mathsf L_s$ is diagonal in the orbit
label, while each $\mathsf L_{so}$ retains the differential coupling along one
trajectory.

The field and passing representations are derived from an $N_\theta$-node
Legendre--Gauss--Lobatto grid after duplicate or boundary-constrained degrees
of freedom are eliminated. For a passing block,
$\dot\tau_{so}=v_{\parallel so}\hat{\mathbf b}\cdot\nabla\theta$, and
$\mathsf T_o$ is the LGL spectral differentiation matrix
\cite{weideman2000differentiation}.
The magnetic geometry, drift, and gyroaverage coefficients are evaluated at
the same nodes. This representation provides spectral convergence for the
smooth field-line structure of local ITG and TEM eigenmodes
\cite{trefethen2000spectral}.

A trapped orbit is parameterized by the periodic bounce coordinate $\tau_b$,
\begin{equation}
  \begin{aligned}
    \theta_o(\tau_b)&=\theta_{c,o}+\theta_{b,o}\sin\tau_b,\\
    v_{\parallel so}(\tau_b)
      &=\operatorname{sgn}(\cos\tau_b)|v_{\parallel so}(\tau_b)|,\\
    v_{\parallel so}\frac{\partial}{\partial\theta}
    &=u_{so}(\tau_b)\frac{\partial}{\partial\tau_b},\\
    u_{so}(\tau_b)&=\frac{|v_{\parallel so}|}
      {\theta_{b,o}|\cos\tau_b|}.
  \end{aligned}
  \label{eq:trapped_bounce_coordinate}
\end{equation}
Here $\theta_{c,o}$ and $\theta_{b,o}$ are the center and half-width of its
accessible interval. The finite bounce-point limit of $u_{so}$ follows from
the local expansion of $B$. The $N_b$ equally spaced nodes in $\tau_b$ and the
corresponding Fourier differentiation matrix $\mathsf T_o$ represent the closed
trajectory, with
$\dot\tau_b=u_{so}(\tau_b)
[\hat{\mathbf b}\cdot\nabla\theta]_{\theta=\theta_o(\tau_b)}$. Field values are
interpolated from the LGL grid to the bounce nodes in barycentric form
\cite{berrut2004barycentric}, and moment deposition uses the
quadrature-weighted adjoint of this interpolation. Along a trapped path,
$J_{\mathrm{loc}}|d\theta_o/d\tau_b|
=(B/B_{\min})|v_{\parallel0}|/u_{so}$ remains
finite at the turning points. Passing deposition uses the LGL path weights,
while trapped deposition uses these regular periodic-path weights.

Interpolation to each orbit and quadrature-weighted deposition back to the
field grid generate the field-coupling and velocity-moment blocks
$\mathsf D_s$ and $\mathsf Q_s$ already introduced in
Eq.~\eqref{eq:discrete_kinetic_field_blocks}.
Thus the kinetic equations are independent at fixed field, and the velocity
moment in quasineutrality provides their only common coupling. For each
species $s$, let $o=1,\ldots,N_{\mathrm{orb},s}$ denote its orbit blocks.
For each orbit,
$\mathsf L_{so}\in\mathbb C^{n_{so}\times n_{so}}$,
$\mathsf Q_{so}\in\mathbb C^{N_\theta\times n_{so}}$, and
$\mathsf D_{so}\in\mathbb C^{n_{so}\times N_\theta}$, while
$\mathsf P\in\mathbb C^{N_\theta\times N_\theta}$.
The orbit-wise differentiation matrix $\mathsf T_o$ is square of size $n_{so}$,
and $\mathsf D_{so}$ maps the $N_\theta$ field coefficients to
$n_{so}$ orbit coefficients.
Passing blocks have $n_{so}$ of order $N_\theta$, trapped blocks have $n_{so}=N_b$,
and the field dimension is $N_\theta$. Increasing the energy or pitch quadrature
order primarily increases $N_{\mathrm{orb},s}$, whereas increasing $N_b$
enlarges the trapped blocks.

Eigenvalues near a complex shift $\zeta$ are obtained from
Eq.~\eqref{eq:generalized_eigenproblem} through
\begin{equation}
  \begin{aligned}
    (\mathsf A-\zeta\mathsf B)^{-1}\mathsf Bx&=\Lambda x,\\
    \omega&=\zeta+\Lambda^{-1}.
  \end{aligned}
  \label{eq:shift_invert_eigenproblem}
\end{equation}
The transformed operator requires repeated solutions of
$(\mathsf A-\zeta\mathsf B)y=r$. Its component equations are
\begin{equation}
  \begin{aligned}
    \mathsf L_{so}^{(\zeta)}y_{so}+\mathsf D_{so}y_\phi&=r_{so},
      &\mathsf L_{so}^{(\zeta)}&=\mathsf L_{so}-\zeta\mathsf I_{so},\\
    \sum_s\sum_{o=1}^{N_{\mathrm{orb},s}}\mathsf Q_{so}y_{so}
      +\mathsf Py_\phi&=r_\phi.&&
  \end{aligned}
  \label{eq:shifted_orbit_block_system}
\end{equation}
With
$\mathsf L=\operatorname{diag}_{s,o}(\mathsf L_{so})$, a shift for which every
$\mathsf L_{so}^{(\zeta)}$ is nonsingular gives directly
\begin{equation}
  (\mathsf L-\zeta\mathsf I)^{-1}
  =\operatorname{diag}_{s,o}\left([\mathsf L_{so}^{(\zeta)}]^{-1}\right).
  \label{eq:shifted_kinetic_direct_sum}
\end{equation}
Each $\mathsf L_{so}^{(\zeta)}$ is factored, the orbit response
$\mathsf R_{so}\in\mathbb C^{n_{so}\times N_\theta}$ is obtained by solving
$\mathsf L_{so}^{(\zeta)}\mathsf R_{so}=\mathsf D_{so}$,
and the field Schur complement is formed as \cite{cottle1974schur}
\begin{equation}
  \mathsf S_\zeta=\mathsf P-
  \sum_s\sum_{o=1}^{N_{\mathrm{orb},s}}\mathsf Q_{so}\mathsf R_{so}.
  \label{eq:field_schur_complement}
\end{equation}
For a valid shift-invert action, $\mathsf S_\zeta$ is nonsingular and is factored once
for the same shift. For each subsequent application of the transformed
operator, the three steps are
\begin{equation}
  \begin{aligned}
    \mathsf L_{so}^{(\zeta)}z_{so}&=r_{so},\\
    \mathsf S_\zeta y_\phi
      &=r_\phi-\sum_s\sum_{o=1}^{N_{\mathrm{orb},s}}\mathsf Q_{so}z_{so},\\
    y_{so}&=z_{so}-\mathsf R_{so}y_\phi.
  \end{aligned}
  \label{eq:orbit_schur_application}
\end{equation}
In the shift-invert application $r=\mathsf Bx$, and the field component $r_\phi$
vanishes by the structure of $\mathsf B$. The orbit inverse actions, field solve, and
back-substitution therefore supply the transformed operator used by the
Arnoldi eigensolver
\cite{ruhe1984rational,saad1980arnoldi,sorensen1992implicit,rommes2008singular}.
A new value of $\zeta$ defines a new setup, while all Arnoldi applications at
the same shift share the prepared orbit responses and field Schur system.

For dense spectral orbit blocks and dense orbit--field maps, preparing one
shift requires
\begin{equation}
  T_{\mathrm{setup}}=O\!\left((N_{\mathrm{orb}}+1)N_\theta^3\right).
  \label{eq:orbit_schur_setup_complexity}
\end{equation}
After this preparation, one application requires
\begin{equation}
  T_{\mathrm{apply}}=O\!\left((N_{\mathrm{orb}}+1)N_\theta^2\right).
  \label{eq:orbit_schur_apply_complexity}
\end{equation}
The shifted solver stores
$O((N_{\mathrm{orb}}+1)N_\theta^2)$ entries, in addition to an Arnoldi
basis of order $O(\ell(N_{\mathrm{orb}}+1)N_\theta)$ for a Krylov dimension $\ell$.
Taking all orbit blocks to have $n_o=O(N_\theta)$, the full unstructured system
has dimension $O((N_{\mathrm{orb}}+1)N_\theta)$. Without orbit decomposition, a
monolithic dense factorization therefore requires
$O((N_{\mathrm{orb}}+1)^3N_\theta^3)$ work in setup and
$O((N_{\mathrm{orb}}+1)^2N_\theta^2)$ work for each inverse application. With
the orbit--Schur decomposition, the corresponding costs are the linear-in-
$N_{\mathrm{orb}}$ scalings in Eqs.~\eqref{eq:orbit_schur_setup_complexity}
and~\eqref{eq:orbit_schur_apply_complexity}. Thus, at fixed field and
per-orbit resolution, the dependence on the number of velocity-space orbits
changes from cubic to linear in setup and from quadratic to linear in each
repeated application. The LGL representation separately reduces the
passing-block and field dimensions required to resolve smooth eigenmodes,
while the Fourier bounce representation controls the trapped-block dimension.
Orbit blocks of equal size are grouped for batched CPU and GPU operations
\cite{abdelfattah2021batched}, which improves execution throughput while
preserving these complexity scalings.

\subsection{Parallel boundary conditions}
\label{sec:parallel_boundary_conditions}

Four boundary conditions are implemented in this code. The \emph{periodic}
boundary condition directly connects the two ends of the domain. It avoids
nonphysical interference from the boundaries when the mode itself is
sufficiently localized, but it does not handle TEMs well. We subsequently
implemented an \emph{open} boundary condition that sets the incoming response
at each boundary to zero while allowing the outgoing response, together with a
boundary condition that places damping buffer regions near the boundaries.
These conditions provide effective treatments for electrostatic kinetic
problems. For the electromagnetic extension described in
Appendix~\ref{app:electromagnetic}, we also implemented an \emph{open-DtN}
boundary condition based on exact non-reflecting
boundary-condition ideas~\cite{keller1989exact}, which provides a
large finite response region outside the core domain to accommodate modes with
potentially long tails.

\section{Electrostatic gyrokinetic mode physics and verification}
\label{sec:results}

The calculations below address three aspects of the linear gyrokinetic
spectrum. We first follow the response of a single ITG branch to its
temperature-gradient drive. We then resolve the competition between ion- and
electron-diamagnetic branches when electrons are treated kinetically. Finally,
we examine the effects of local Miller geometry and of the field-line domain
required to resolve an extended mode. Branch identity is assessed jointly from the
continuity of the complex frequency and the evolution of the parallel mode
structure; independent calculations provide the corresponding numerical
comparisons.

The normal-mode convention of Sec.~\ref{sec:continuous_normal_modes} is used
throughout: $\omega_r<0$ denotes propagation in the ion diamagnetic direction,
whereas $\omega_r>0$ denotes propagation in the electron diamagnetic direction.
External frequencies are converted to the sign convention of MGK before comparison.
Because the reference calculations use different length and thermal-speed
conventions, the frequency normalization is stated for each case.

\subsection{\texorpdfstring{Ion-temperature-gradient\protect\\
dependence in $s$--$\alpha$ geometry}%
{Ion-temperature-gradient dependence in s-alpha geometry}}

We first consider how the adiabatic-electron ITG branch responds to increasing
temperature-gradient drive. The calculation uses kinetic ions restricted to
passing orbits in circular $s$--$\alpha$ geometry, with $q=\hat{s}=1$,
$\alpha=\theta_0=0$, $r/R_0=0.05$, $L_n/R_0=0.25$, $T_e/T_i=1$, and
$k_y\rho_i=0.45/\sqrt{2}$. At fixed density gradient, perpendicular scale, and
magnetic geometry, $\eta_i=L_n/L_{T_i}$ is varied from $2.3$ to $2.7$. Since
$R_0/L_{T_i}=\eta_iR_0/L_n$, this scan changes the temperature-gradient
contribution to $\omega_{\ast i}^{T}$ in Eq.~\eqref{eq:implemented_gk} while
holding the other drives fixed.

\begin{figure*}[t]
  \centering
  \includegraphics[width=0.98\textwidth]{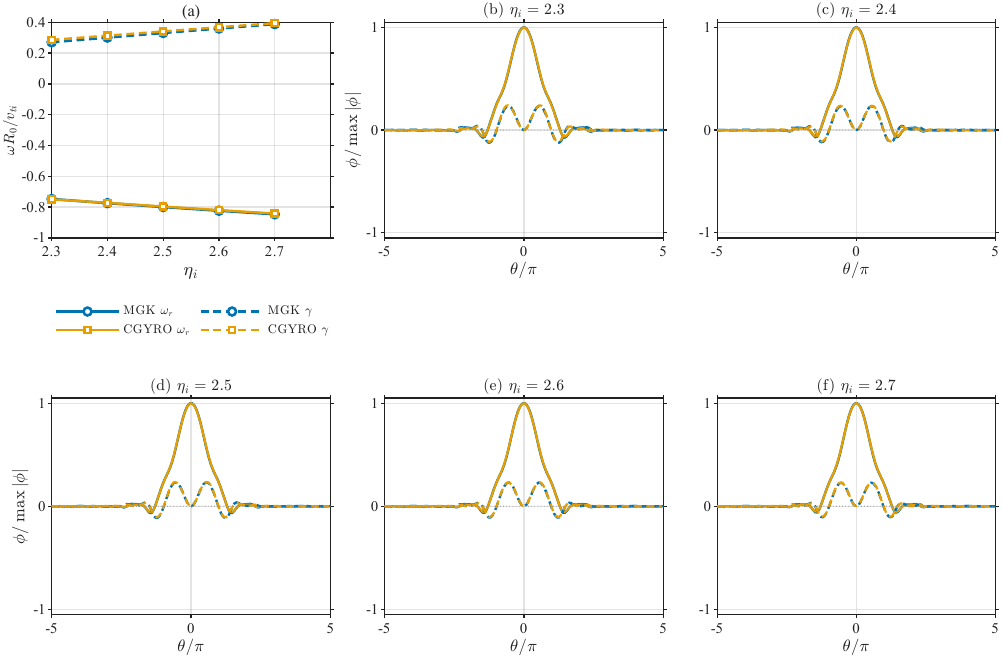}
  \caption{Adiabatic-electron ITG dependence on $\eta_i$ in circular
  $s$--$\alpha$ geometry. Panel (a) gives the MGK and CGYRO real frequencies
  and growth rates in units of $v_{\mathrm{th},i}/R_0$. Panels (b)--(f) give the
  phase-aligned complex electrostatic potentials at the five values of
  $\eta_i$, with each potential normalized by its maximum magnitude. Solid and
  dashed curves denote the real and imaginary components, respectively. Blue
  denotes MGK and orange denotes CGYRO; circles and squares distinguish them in
  panel (a).}
  \label{fig:salpha_itg_eta_scan}
\end{figure*}

The growth rate and $|\omega_r|$ increase monotonically across the scan, while
the potential retains a ballooning structure that evolves continuously with
$\eta_i$ [Fig.~\ref{fig:salpha_itg_eta_scan}]. The continuity of both the
complex frequency and the parallel structure identifies a single
ion-diamagnetic ITG branch. Because the other gradients and geometric
coefficients are fixed, this trend is consistent with the increasing
energy-dependent temperature-gradient drive in $\omega_{\ast i}^{T}$.

MGK and CGYRO trace this branch with a maximum relative complex-frequency
difference below $2\%$. After pairwise phase alignment and peak normalization,
the real and imaginary parts exhibit matching ballooning envelopes and parallel
phase variation at every value of $\eta_i$.

\subsection{Kinetic-electron ITG--TEM competition in the Cyclone Base Case}

Treating electrons kinetically introduces electron-diamagnetic modes into the
spectral range occupied by the ITG branch. We examine the resulting branch
competition at the Cyclone Base Case parameters used in
Refs.~\cite{dimits2000comparisons,rewoldt2007linear}: $q=1.4$,
$\hat{s}=0.776$, $r/R_0=0.18$, $R_0/L_n=2.22$,
$R_0/L_{Ti}=R_0/L_{Te}=6.92$, $T_i=T_e$, and the physical electron-to-ion
mass ratio. For this singly charged, equal-temperature case,
$c_s=v_{\mathrm{th},i}$, $\rho_s=\rho_i$, and $k_y=k_\theta$ at the reference
surface. Both species include passing and trapped populations. Passing
trajectories satisfy the open field-line boundary condition, whereas trapped
trajectories close at their bounce points. The ITG and TEM roots are continued
separately in $k_y\rho_s$, and their growth rates are then compared to determine
the dominant mode.

\begin{figure*}[t]
  \centering
  \includegraphics[width=0.98\textwidth]{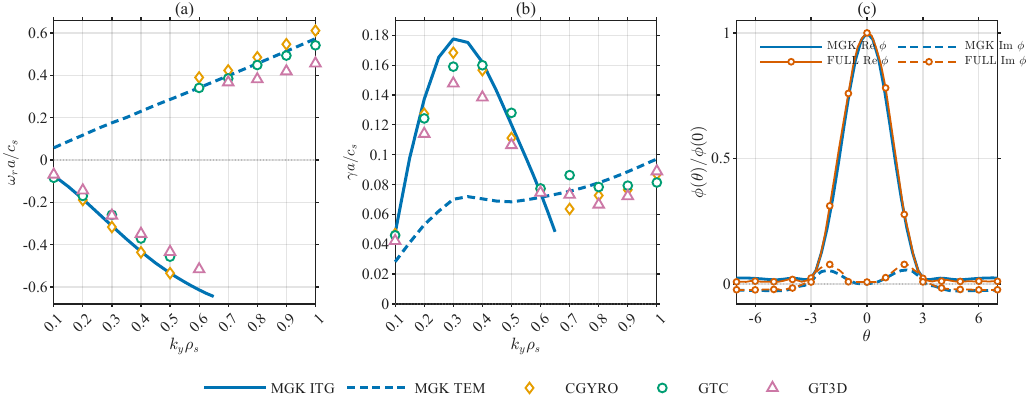}
  \caption{Kinetic-electron ITG--TEM competition at the Cyclone Base Case
  parameters of Rewoldt et al. Panels (a) and (b) give the real frequencies
  and growth rates for $R_0/L_{Ti}=R_0/L_{Te}=6.92$ and $R_0/L_n=2.22$.
  Blue solid and dashed curves are the separately continued MGK ITG and TEM
  branches; orange diamonds are
  the mode-matched CGYRO results, while circles and triangles are the
  dominant-mode GTC and GT3D results digitized from Fig.~2 of
  Ref.~\citenum{rewoldt2007linear}. Panel (c) gives the normalized potential
  eigenfunction for the companion TEM case with $R_0/L_{Ti}=2.22$
  ($\eta_i=1$) and $k_\theta\rho_i=0.335$, with the remaining parameters
  unchanged. In panel (c), blue denotes MGK and vermilion circles denote FULL;
  solid and dashed curves denote the real and imaginary parts, respectively.
  FULL data are digitized from Fig.~5(a) of Ref.~\citenum{rewoldt2007linear}. The global phase
  is chosen so that $\phi(0)$ is real and positive. Frequencies are expressed
  in units of $c_s/a$.}
  \label{fig:cbc_itg_tem_branches}
\end{figure*}

The separately continued roots reveal two distinct branches
[Fig.~\ref{fig:cbc_itg_tem_branches}(a),(b)]. The negative-frequency ITG growth
rate initially rises with $k_y\rho_s$ and then decreases, whereas the
positive-frequency TEM growth rate varies more weakly over the
shorter-wavelength part of the scan. Their growth rates cross near
$k_y\rho_s\simeq0.6$, changing the dominant mode from ITG to TEM. Each
eigenvalue remains continuous through the crossing. Thus, the sign reversal of
the real frequency selected by a dominant-mode scan reflects a switch between
branches, not a reversal of propagation along a single branch.

The shorter-wavelength reduction of the ITG growth rate is consistent with
stronger ion-FLR averaging. The persistence of the TEM branch over the same
interval is consistent with the trapped-electron magnetic-drift response to the
electron gradients. The mode-matched CGYRO curve and the dominant-mode GTC and
GT3D data digitized from Ref.~\citenum{rewoldt2007linear} exhibit qualitatively
similar wavenumber dependence. Although the peak growth rates and transition
locations differ, all four calculations have an ITG-dominated low-$k_y\rho_s$
interval and a TEM-dominated high-$k_y\rho_s$ interval.

The parallel mode structure of the electron-diamagnetic branch is examined with
the companion $\eta_i=1$ case of Ref.~\citenum{rewoldt2007linear}. This case
retains the magnetic geometry, density gradient, and electron-temperature
gradient of the wavenumber scan while setting $R_0/L_{Ti}=2.22$ and
$k_\theta\rho_i=0.335$. The reduced ion-temperature gradient places the ITG
branch below threshold and leaves the TEM as the unstable branch. After fixing
the phase and normalizing by $\phi(0)$, the MGK potential matches the central
localization and symmetric imaginary side lobes of the FULL eigenfunction
digitized from the reference [Fig.~\ref{fig:cbc_itg_tem_branches}(c)].

As a strong-gradient extension of the CBC calculation, we retain its magnetic
geometry, $T_i=T_e$, and the physical electron-to-ion mass ratio, and reduce
the density-gradient scale to $L_n/R_0=0.018$. Setting $\eta_e=3.13$ then gives
$R_0/L_{T_e}\simeq174$, while $\eta_i=0$ suppresses the
ion-temperature-gradient drive. We examine the resulting kinetic-electron mode
at $k_y\rho_i=0.7$.

\begin{figure}[pos=t]
  \centering
  \includegraphics[width=\columnwidth]
    {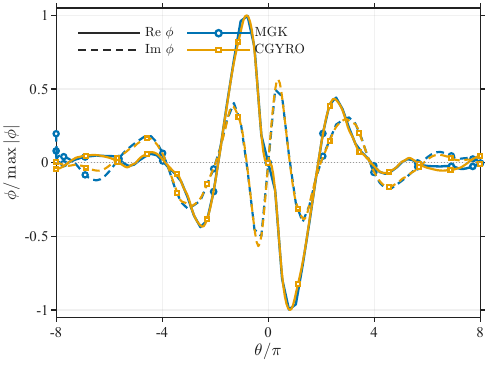}
  \caption{Strong-gradient kinetic-electron eigenfunction in circular
  $s$--$\alpha$ geometry for $q=1.4$, $\hat{s}=0.776$, $r/R_0=0.18$,
  $L_n/R_0=0.018$, $\eta_i=0$, $\eta_e=3.13$, and $k_y\rho_i=0.7$.
  Blue circles and orange squares denote MGK and CGYRO, respectively; solid
  and dashed lines denote the real and imaginary parts of $\phi$. The profiles
  are phase aligned and normalized independently by $\max_\theta|\phi|$ over
  $-8\pi\leq\theta\leq8\pi$.}
  \label{fig:strong_gradient_tem_mode}
\end{figure}

The resulting positive-frequency mode has several parallel nodes, with an
approximately odd dominant real component
[Fig.~\ref{fig:strong_gradient_tem_mode}]. MGK and CGYRO give
$\omega R_0/v_{\mathrm{th},i}\simeq23.3+13.8\iu$ and $23.4+13.8\iu$, respectively, and
their complex frequencies differ by about $0.3\%$. The phase-aligned
potentials reproduce the same sequence of central oscillations and extended
parallel tails.

\subsection{Electrostatic modes in Miller geometry}

We next consider how the kinetic response changes when the analytic
$s$--$\alpha$ coefficients are replaced by those of a local Miller equilibrium.
The kinetic equation, orbit classification, field closure, and eigensolver are
unchanged; only the magnetic-field, metric, and drift coefficients are supplied
by the Miller representation. For both circular and shaped surfaces, the
coefficients are evaluated with the GACODE \code{geo}
conventions~\cite{candy2016cgyro}.

\subsubsection{Triangularity dependence of the ITG mode}

An adiabatic-electron Miller case using the GACODE input
conventions~\cite{candy2016cgyro} is used
to probe the response of the ITG branch to triangular shaping. At fixed
$k_y\rho_s=0.3$, $\delta$ is varied over $-0.4,-0.2,0,0.2,$ and $0.4$. The
remaining parameters are $R_0/a=3$, $r/a=0.5$, $q=2$, $\hat{s}=1$,
$a/L_n=1$, $a/L_{Ti}=3$, $T_i=T_e$, $\kappa=1$, and
$s_\kappa=s_\delta=\Delta'=\alpha=0$. The negative-frequency ITG root is
continued from the circular surface separately toward positive and negative
triangularity.

\begin{figure*}[t]
  \centering
  \includegraphics[width=0.98\textwidth]{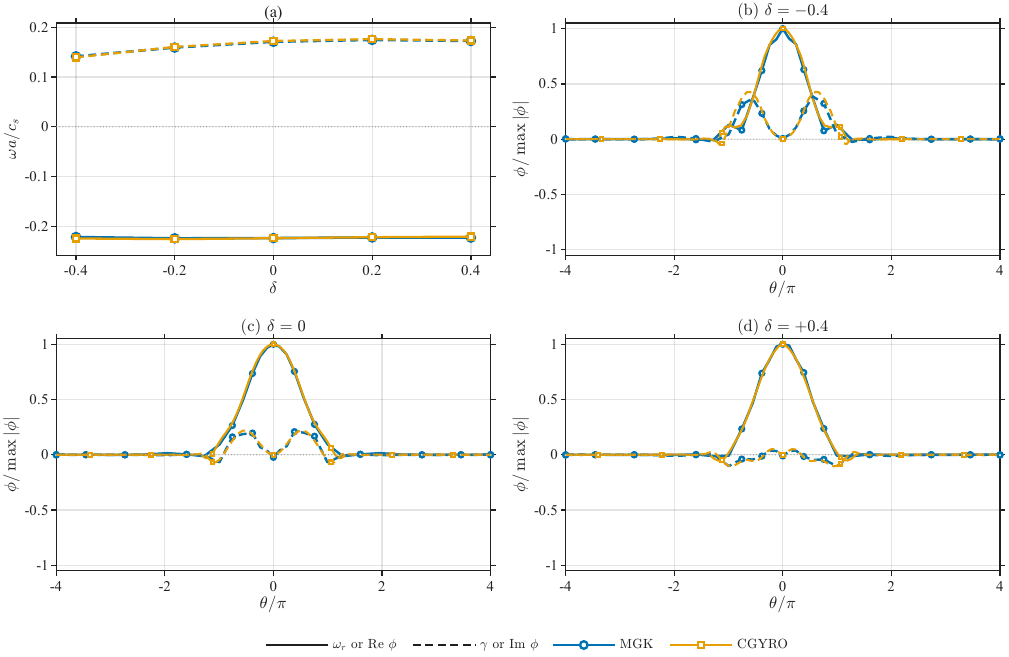}
  \caption{Triangularity dependence of the adiabatic-electron ITG mode in
  Miller geometry. Panel (a) gives the MGK and CGYRO real frequencies and
  growth rates at fixed $k_y\rho_s=0.3$, expressed in units of $c_s/a$.
  Panels (b)--(d) give the phase-aligned complex potentials at
  $\delta=-0.4$, $0$, and $0.4$, with each potential normalized by its maximum
  magnitude. Circles denote MGK and squares denote CGYRO. Blue denotes MGK and
  orange denotes CGYRO; solid curves
  denote $\omega_r$ in panel (a) and $\operatorname{Re}\phi$ in panels
  (b)--(d), while dashed curves denote $\gamma$ and
  $\operatorname{Im}\phi$, respectively.}
  \label{fig:miller_itg_triangularity}
\end{figure*}

The growth rate rises from negative triangularity to a broad maximum near
$\delta=0.2$ and then varies only weakly at larger positive triangularity,
whereas $\omega_r$ remains close to $-0.22\,c_s/a$
[Fig.~\ref{fig:miller_itg_triangularity}]. The potential becomes increasingly
localized about the outboard midplane but evolves smoothly throughout the
scan, identifying a continuous ion-diamagnetic ITG branch.

Triangularity changes both the magnetic-drift projection and the metric factors
entering $k_\perp(\theta)$ over the outboard region sampled by the mode. The rise
in growth rate toward $\delta=0.2$ is consistent with a stronger
unfavorable-curvature response and increased outboard localization. The
subsequent leveling is consistent with stronger off-midplane FLR attenuation as
$k_\perp(\theta)$ increases and the parallel structure narrows.

MGK and CGYRO independently follow this continuous branch, with a maximum
relative complex-frequency difference below $1.5\%$. At the three displayed
values, their phase-aligned potentials have matching ballooning envelopes and
parallel phase variation.

\subsubsection{Kinetic-electron TEM and parallel extent}

The kinetic-electron Rewoldt TEM~\cite{rewoldt2007linear} raises a distinct
numerical issue because its potential extends over several poloidal turns.
On a circular Miller surface, the mode couples trapped-electron magnetic-drift
and bounce dynamics to passing-particle streaming. Both passing and trapped
trajectories are therefore retained, and the computed frequency and mode
structure can depend on the represented field-line length.

\begin{figure*}[t]
  \centering
  \includegraphics[width=0.98\textwidth]{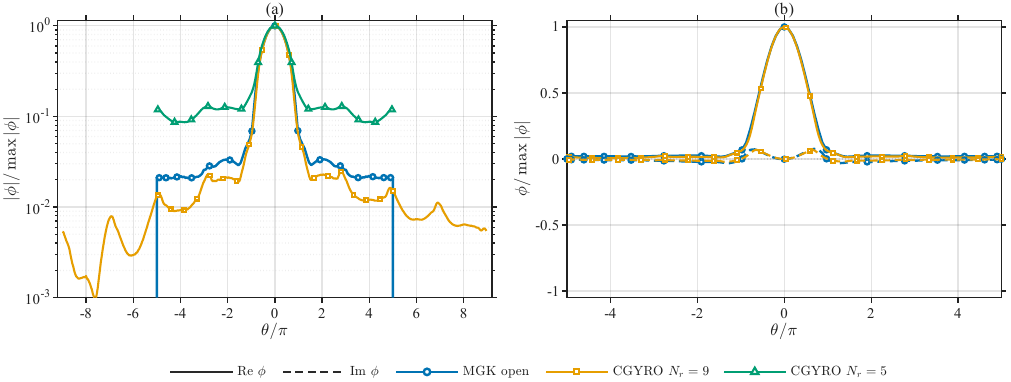}
  \caption{Parallel-extent comparison for the kinetic-electron TEM in circular
  Miller geometry. Panel (a) gives the normalized potential magnitude over the
  available ballooning domains. Here $N_r$ is the number of CGYRO radial
  harmonics: the $N_r=5$ and $N_r=9$ calculations give
  $\omega a/c_s=0.2000+0.2535\iu$ and $0.1915+0.2047\iu$, with normalized
  outer-turn maxima of $0.119$ and $6.4\times10^{-3}$, respectively.
  Panel (b) compares the phase-aligned real and imaginary parts of the
  $N_r=9$ CGYRO potential with the open-boundary MGK result. Blue circles
  denote MGK, orange squares denote CGYRO $N_r=9$, and green triangles denote
  CGYRO $N_r=5$; solid and dashed curves in panel (b) denote the real and
  imaginary parts, respectively. The MGK frequency is
  $\omega a/c_s=0.1870+0.2100\iu$. Relative to CGYRO $N_r=9$, the
  complex-frequency difference is about $2.5\%$ and the potential overlap is
  $0.9991$. All potentials are normalized by their maximum magnitude.}
  \label{fig:tem_domain}
\end{figure*}

Extending the CGYRO domain from five to nine radial harmonics strongly
suppresses the outer-turn amplitude and shifts the complex frequency
[Fig.~\ref{fig:tem_domain}(a)], identifying the shorter-domain result as
boundary sensitive. The much smaller tail at $N_r=9$ motivates its use as the
extended-domain reference.

In the MGK representation, the open boundary condition prescribes the incoming
passing response while allowing the outgoing response to follow from the orbit
equation; trapped trajectories close at their bounce points. The MGK frequency
and phase-aligned potential agree closely with the extended-domain CGYRO result
over the central five poloidal turns [Fig.~\ref{fig:tem_domain}(b)]. This
agreement in frequency and spatial structure indicates that the open
representation resolves the extended parallel structure in this case.

\section{Electrostatic computational performance}
\label{sec:performance}

The physical comparisons in Sec.~\ref{sec:results} identify the branches and
compare their frequencies and eigenfunctions. We first measure the end-to-end
cost of following the electrostatic branches through parameter scans. Two
workloads are used: a five-point adiabatic-electron
$\eta_i$ scan and a 29-point kinetic-electron ITG--TEM scan in $k_y\rho_s$.
The latter uses rounded CBC parameters, $\hat{s}=0.8$, $R_0/L_n=2.2$, and
$R_0/L_{Ti}=R_0/L_{Te}=6.9$.

The first point of each workload provides the branch anchor. At every subsequent
parameter value, the previously converged mode supplies only spectral initial
information; the orbit matrices, shifted factorization, and eigensolution are
recomputed for the new parameter value. The reported averages exclude the
initial anchor and therefore measure changed-point continuation throughput
rather than the latency of an isolated cold-start calculation.

\subsection{Resolution-dependent scan cost}

\begin{figure*}[t]
  \centering
  \includegraphics[width=0.98\textwidth]{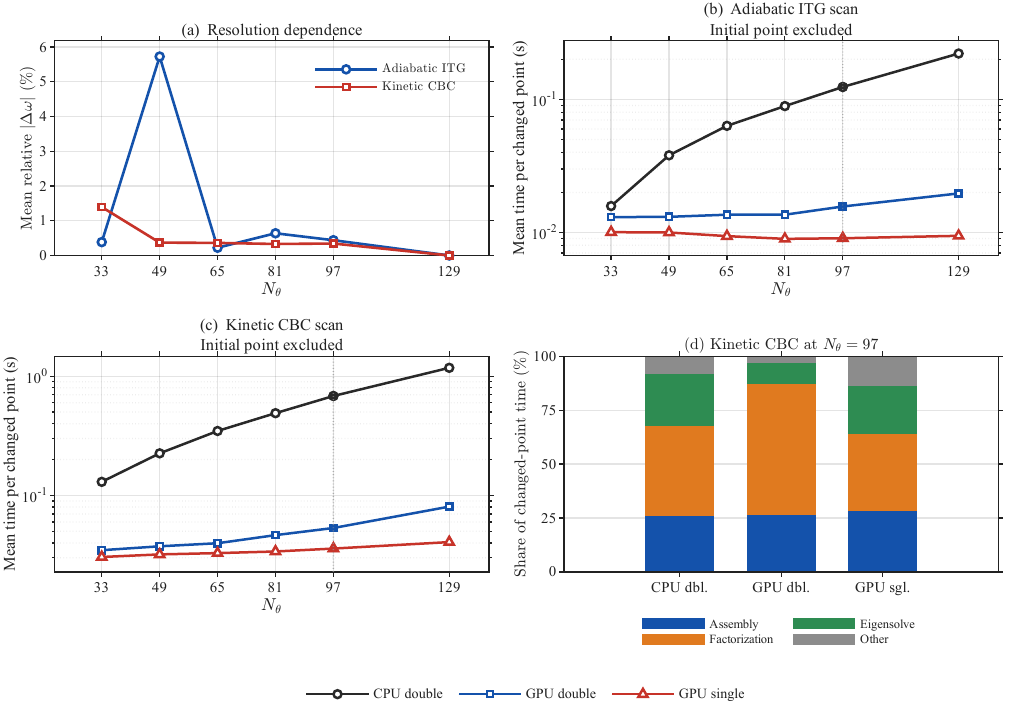}
  \caption{Parallel-resolution dependence and end-to-end changed-point
  throughput. Panel (a) gives the scan-averaged relative complex-frequency
  difference from the highest-resolution internal reference, defined by the
  CPU double-precision $N_\theta=129$ result, for the adiabatic-electron
  $\eta_i$ and kinetic-electron ITG--TEM workloads. With
  $0\leq E_s/T_s\leq12.5$, the adiabatic-electron workload uses 16 energy and
  16 pitch quadrature points, whereas the kinetic-electron workload uses 16
  energy and 24 pitch quadrature points for each kinetic species, together with
  $N_b=24$ bounce points for trapped-orbit blocks. Panels (b) and (c) give the
  mean wall time per changed point after excluding the initial workload anchor;
  the vertical guide marks $N_\theta=97$. Panel (d) decomposes the corresponding
  kinetic-electron time into assembly, factorization, eigensolution, and
  remaining solver overhead.}
  \label{fig:electrostatic_scan_performance}
\end{figure*}

The parallel resolution is varied over $N_\theta=33$--129 with a fixed velocity
grid for each workload. We quantify the frequency variation by the scan average
$\overline{\epsilon}_\omega(N_\theta)=N_p^{-1}\sum_j
|\omega_j(N_\theta)-\omega_j^{\mathrm{ref}}|/
|\omega_j^{\mathrm{ref}}|$, where matching parameter values and branches are
compared and the CPU double-precision result at $N_\theta=129$ defines
$\omega_j^{\mathrm{ref}}$. This is the highest-resolution internal reference
for measuring parallel-discretization variation; independent physical
comparisons are reported in Sec.~\ref{sec:results}. The scan-averaged difference
remains below $1\%$ for both workloads throughout $N_\theta\geq65$
[Fig.~\ref{fig:electrostatic_scan_performance}(a)]. We use $N_\theta=97$ as a
common operating point for the backend comparison while retaining
$N_\theta=129$ as the internal resolution reference.

Increasing $N_\theta$ enlarges every orbit block and the field Schur system. The
kinetic-electron workload also contains passing and trapped populations over a
denser set of velocity-space orbits, which contributes to its larger cost.

\subsection{CPU and GPU scan throughput}

The measurements in Fig.~\ref{fig:electrostatic_scan_performance}(b)--(d) were
obtained on a system equipped with an AMD EPYC 9354 32-Core Processor and an
NVIDIA GeForce RTX 5070 Ti GPU. At $N_\theta=97$, the adiabatic-electron workload
requires about $0.124$~s per changed point on the double-precision CPU,
$0.0157$~s on the double-precision GPU, and $0.00905$~s on the single-precision GPU.
The corresponding kinetic-electron costs are about $0.683$, $0.0533$, and
$0.0358$~s per changed point.

For the kinetic-electron workload, the double- and single-precision GPU paths
reduce the mean changed-point time by factors of about $12.8$ and $19.0$,
respectively, relative to the double-precision CPU path. The CPU and GPU
double-precision backends apply the same discretized problem and produce
matching frequencies. For both workloads at $N_\theta=97$, the frequency
difference between the single- and double-precision GPU results is smaller than
the parallel-resolution change from $N_\theta=97$ to 129. Assembly,
factorization, and eigensolution all contribute to the end-to-end GPU gain
[Fig.~\ref{fig:electrostatic_scan_performance}(d)].

The grids used for the physics figures provide complementary changed-point
measurements. After their initial anchors, the mapped-grid CPU calculations
average about $0.64$~s per point for the CBC branch scan and $0.88$~s per point
for the Miller triangularity scan. These scans use the physical grids selected
for the corresponding mode calculations together with the orbit-factorized
eigensolver.

\section{Conclusions}
\label{sec:conclusions}

In this work, we extended the MGK code to kinetic electrons, the full
$s$--$\alpha$ geometry, and Miller geometry, and optimized its computational
speed. The main method is to construct the velocity-space discretization using
orbit invariants and then apply Schur-complement elimination, thereby reducing
the complexity of the eigenvalue problem. The implementation was validated
against CGYRO and other codes for electrostatic ITG cases with adiabatic
electrons, kinetic-electron CBC cases, ITG and TEM cases in Miller geometry, achieving a speedup of three orders of magnitude over CGYRO for the same
problems. In future work, we will attempt to extend this approach to two-dimensional eigenvalue and initial-value codes, and to address nonlinear simulations and transport problems without compromising the numerical structure~\cite{falessi2019transport}.

\section*{Acknowledgments}
The authors thank Xiang Jian, Jiaqi Dong, Huishan Cai and Chang Liu for helpful discussions.

\section*{Code availability}
The MGK source code used in this work is openly available at
\url{https://github.com/FusionAlpha/mgk}.

\appendix

\section{Electromagnetic extension}
\label{app:electromagnetic}

\subsection{Normalized electromagnetic model}
\label{app:em_model}

The electrostatic formulation in Secs.~\ref{sec:model} and
\ref{sec:numerical_formulation} is extended by retaining the parallel vector
potential and the parallel magnetic perturbation in the gyroaveraged
generalized potential~\cite{brizard2007foundations,candy2016cgyro}. All
quantities below use the normalization of Sec.~\ref{sec:normalized_eigenproblem}.
In particular, $A_\parallel$ is normalized by
$T_i/(Z_i e v_{\mathrm{th},i})$, $\delta B_\parallel$ by $B_0$, and
\begin{equation}
  \beta_e=\frac{2\mu_0n_eT_e}{B_0^2}.
  \label{eq:em_beta_definition}
\end{equation}
The velocities $v_\parallel$ and $v_\perp$ in the field-coupling factors are
normalized by the thermal speed of species $s$. Define
$u_s=v_{\mathrm{th},s}/v_{\mathrm{th},i}$, $\widehat B=B/B_0$, and
\begin{equation}
  \boldsymbol{\Psi}=
  \begin{pmatrix}
    \phi & A_\parallel & \delta B_\parallel
  \end{pmatrix}^{T},
  \qquad
  \mathbf J_s=
  \begin{pmatrix}
    J_0(a_s)\\[2pt]
    -u_sv_\parallel J_0(a_s)\\[2pt]
    \displaystyle
    \frac{\tau_s}{z_s}\frac{v_\perp^2}{2\widehat B}
    \frac{2J_1(a_s)}{a_s}
  \end{pmatrix}.
  \label{eq:em_field_coupling_factors}
\end{equation}
The finite-Larmor-radius factor in the third component is evaluated with its
regular limit $2J_1(a_s)/a_s\to1$ as $a_s\to0$. The generalized potential
sampled by species $s$ is
\begin{equation}
  \chi_s=\mathbf J_s^{T}\boldsymbol{\Psi}.
  \label{eq:em_generalized_potential}
\end{equation}

With this replacement, Eq.~\eqref{eq:implemented_gk} generalizes to
\begin{equation}
  \begin{aligned}
    &-\mathrm{i}\frac{R_0}{v_{\mathrm{th},i}}
      \left[v_\parallel\nabla_\parallel
      +\mathrm{i}\mathbf k_\perp\cdot\mathbf v_{ds}\right]
      \left[g_s+\frac{z_s}{\tau_s}F_{0s}\chi_s\right]\\
    &\qquad
      -\frac{z_s}{\tau_s}F_{0s}
      \widehat\omega_{\ast s}^{T}\chi_s
      =\omega g_s.
  \end{aligned}
  \label{eq:implemented_em_gk}
\end{equation}
Thus the orbit operator acting on $g_s$ is unchanged; the additional fields
enter only through the field-driven response on each independent orbit.

The three field equations can be written in the same normalized variables as
\begin{equation}
  \begin{aligned}
    &\sum_{s=1}^{N_{\mathrm{kin}}}z_s\bar n_s
      \int d^3v\,\mathbf J_sg_s\\
    &\quad+
      \left[
        \mathbf C_0+
        \sum_{s=1}^{N_{\mathrm{kin}}}
        \frac{\bar n_sz_s^2}{\tau_s}
        \int d^3v\,F_{0s}\mathbf J_s\mathbf J_s^{T}
      \right]\boldsymbol{\Psi}=0,
  \end{aligned}
  \label{eq:implemented_em_field_closure}
\end{equation}
where
\begin{equation}
  \mathbf C_0=
  \operatorname{diag}\!\left(
    -\sum_{s=1}^{N_{\mathrm{sp}}}
      \frac{\bar n_sz_s^2}{\tau_s},
    \frac{2k_\perp^2}{\beta_e},
    \frac{2}{\beta_e}
  \right).
  \label{eq:em_vacuum_field_block}
\end{equation}
The first, second, and third rows of
Eq.~\eqref{eq:implemented_em_field_closure} are quasineutrality, parallel
Amp\`ere's law, and perpendicular pressure balance, respectively. The
two-field model is obtained by deleting the third component and the
corresponding row and column. Retaining only the first component recovers
Eq.~\eqref{eq:implemented_quasineutrality} exactly. After orbit discretization,
the same Schur reduction described in
Sec.~\ref{sec:phase_space_discretization} applies with a larger field block.

\subsection{Electromagnetic verification}
\label{app:em_validation}

\subsubsection{ITG--TEM--KBM transition in the Cyclone Base Case}
\label{app:em_xie}

The electromagnetic implementation is first tested against the Cyclone Base
Case scan of Xie et al.~\cite{xie2016sensitivity}, which contains drift-wave
and kinetic-ballooning branches within a single parameter sequence. The
equilibrium and gradient parameters are $q=1.4$, $\hat{s}=0.78$,
$r/R_0=0.18$, $R_0/L_n=2.2$, and
$R_0/L_{Ti}=R_0/L_{Te}=6.9$, with $T_i=T_e$, the physical electron--ion mass
ratio, and $k_\theta\rho_i=0.22$. Ions and electrons are kinetic, the magnetic
mirror force is retained, and the two-field model evolves $\phi$ and
$A_\parallel$ while neglecting $\delta B_\parallel$. The MGK calculation uses
the open domain $\theta\in[-10\pi,10\pi]$ with $N_\theta=161$ and the velocity
grid $E_{\max}/T_s=12.5$, $N_E=16$, $N_{\mathrm{pitch}}=24$, and $N_b=48$.
The electron beta is scanned in increments of $0.02\%$.

\begin{figure*}[t]
  \centering
  \includegraphics[width=0.98\textwidth]{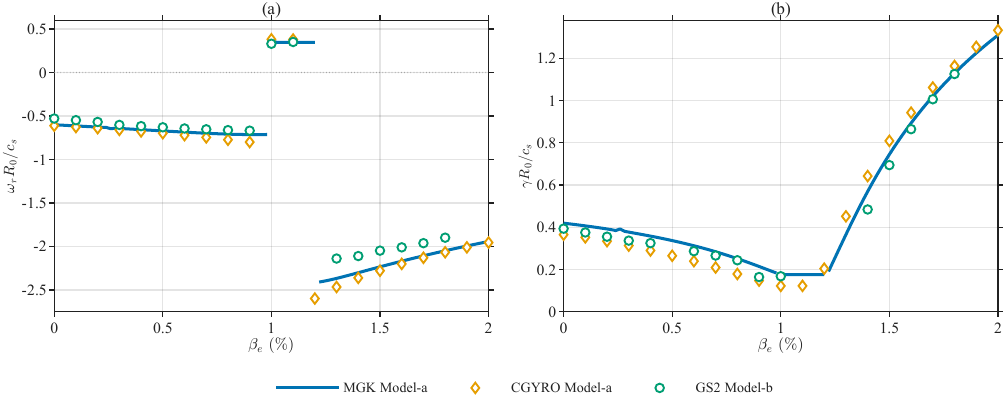}
  \caption{Electromagnetic Cyclone Base Case scan. Panels (a) and (b) show
  $\omega_rR_0/c_s$ and $\gamma R_0/c_s$, respectively, as functions of
  $\beta_e$ in percent. The blue solid curves are the dominant MGK Model-a
  results, interrupted at branch changes; orange diamonds are the
  corresponding CGYRO Model-a results. Green circles are GS2 Model-b values
  digitized from Fig.~1 of Xie et al.~\cite{xie2016sensitivity} and are shown
  as a literature reference rather than as a same-model pointwise comparison.
  The MGK and CGYRO calculations use kinetic ions and electrons and the
  two-field system $\phi+A_\parallel$, without $\delta B_\parallel$.}
  \label{fig:xie2016_electromagnetic_cbc}
\end{figure*}

At low $\beta_e$, the dominant solution is the ion-diamagnetic ITG branch,
whose growth rate decreases as electromagnetic effects increase
[Fig.~\ref{fig:xie2016_electromagnetic_cbc}]. The dominant mode changes to the
electron-diamagnetic TEM between $\beta_e=0.98\%$ and $1.00\%$, and then to the
rapidly growing KBM between $1.20\%$ and $1.22\%$. The breaks in the MGK
frequency curve mark these changes of eigenmode and do not represent
discontinuities of an individually continued branch.

MGK and the same-model CGYRO calculation recover the same ordering and beta
dependence of the three branches. The Model-b GS2 points provide additional
context for the location of the electromagnetic transition, but their
quantitative displacement from the Model-a curves is expected because the
equilibrium implementations differ. Reproducing both the low-beta drift-wave
response and the onset of the shear-Alfv\'enic KBM in the same scan provides an
implementation-level validation of the kinetic-species coupling through
$A_\parallel$.

\subsubsection{Large-aspect-ratio electromagnetic KBM benchmark}
\label{app:em_shen}

A second test isolates the electromagnetic field coupling in the
large-aspect-ratio KBM benchmark of Shen et al.~\cite{shen2025electromagnetic}.
The parameters are $r/R_0=0.0018$, $q=2$, $\hat{s}=1$,
$k_\theta\rho_i=0.3$, and
$R_0/L_n=R_0/L_{Ti}=R_0/L_{Te}=5$, with kinetic electrons, the physical mass
ratio, and the magnetic mirror force retained. In the $s$--$\alpha$ geometry,
$\alpha=80\beta_e$, so the pressure-gradient drive changes consistently as
$\beta_e$ is varied from $0.8\%$ to $2.5\%$. The periodic parallel domain is
$\theta\in[-5\pi,5\pi]$ with $N_\theta=241$; the velocity grid uses
$E_{\max}/T_s=12.5$, $N_E=16$, $N_{\mathrm{pitch}}=24$, and $N_b=24$.
Two calculations are compared: a two-field model with $\phi$ and $A_\parallel$
and a three-field model that also retains $\delta B_\parallel$.

\begin{figure*}[t]
  \centering
  \includegraphics[width=0.98\textwidth]{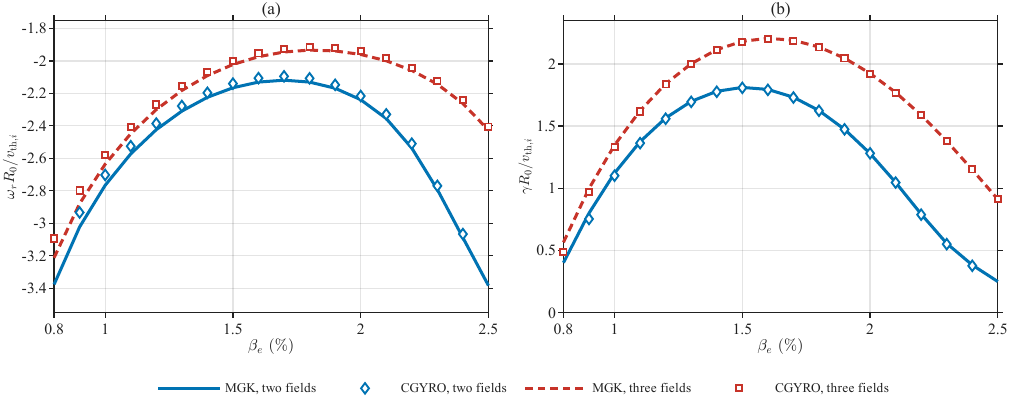}
  \caption{Large-aspect-ratio electromagnetic KBM benchmark based on the
  parameters of Shen et al.~\cite{shen2025electromagnetic}. Panels (a) and (b)
  show $\omega_rR_0/c_s$ and $\gamma R_0/c_s$, respectively, over the
  $\beta_e$ scan. Blue solid curves and blue diamonds denote the MGK and CGYRO
  two-field ($\phi+A_\parallel$) results. Red dashed curves and red squares
  denote the corresponding three-field
  ($\phi+A_\parallel+\delta B_\parallel$) results. Both codes use the matched
  periodic parallel domain and the same $s$--$\alpha$ pressure-gradient
  prescription, $\alpha=80\beta_e$.}
  \label{fig:shen2025_electromagnetic_kbm}
\end{figure*}

Both field models exhibit the same KBM evolution: the growth rate rises rapidly
from the low-beta onset and then changes more gradually, while the magnitude of
the ion-diamagnetic real frequency decreases across the scan
[Fig.~\ref{fig:shen2025_electromagnetic_kbm}]. Retaining
$\delta B_\parallel$ increases the growth rate and shifts the real frequency at
each beta, demonstrating a finite compressional-magnetic response rather than
a relabeling of the two-field eigenvalue.

Over the converged scan points, the mean relative MGK--CGYRO difference in the
complex frequency is $1.23\%$ for the two-field model and $1.20\%$ for the
three-field model; the corresponding maximum differences are $3.32\%$ and
$4.52\%$, respectively, and occur near the low-beta branch onset. Thus MGK
reproduces not only the $A_\parallel$-mediated KBM branch but also the systematic
change caused by $\delta B_\parallel$. The agreement of both field variants
with an independent code supports the implementation of the full
electromagnetic field coupling.

\begin{figure*}[t]
  \centering
  \includegraphics[width=0.98\textwidth]{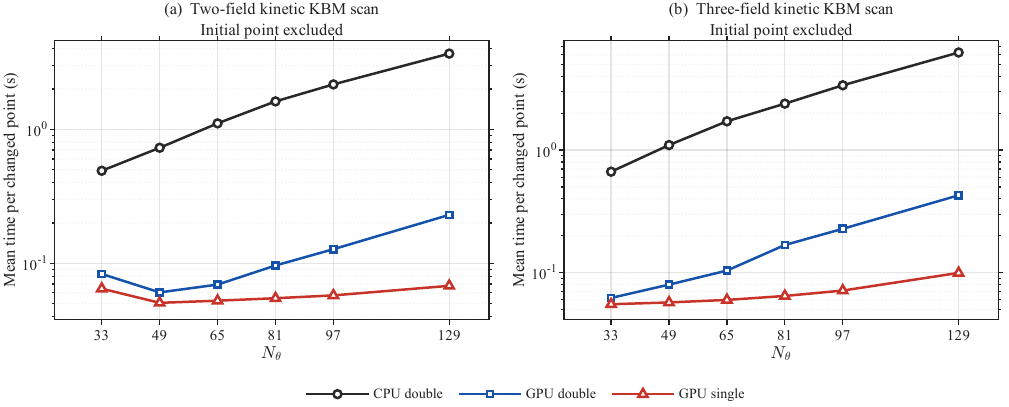}
  \caption{End-to-end changed-point throughput for the large-aspect-ratio Shen
  electromagnetic KBM scan. Panels (a) and (b) show the mean wall time per
  changed point for the two-field ($\phi+A_\parallel$) and three-field
  ($\phi+A_\parallel+\delta B_\parallel$) systems, respectively. Each mean is
  taken over the four non-anchor points of the five-point $\beta_e$ scan; the
  initial $\beta_e=2.0\%$ branch anchor is excluded. Black circles, blue
  squares, and red triangles denote the CPU double-precision, GPU
  double-precision, and GPU single-precision backends. Both panels use kinetic
  ions and electrons, the same velocity-space grid, and the hardware specified
  for Fig.~\ref{fig:electrostatic_scan_performance}.}
  \label{fig:shen_electromagnetic_scan_performance}
\end{figure*}

\subsection{Electromagnetic scan throughput}
\label{app:em_performance}

The backend comparison is extended to the large-aspect-ratio electromagnetic
KBM case in Sec.~\ref{app:em_shen}. For each field model, a five-point scan uses
$\beta_e=1.8\%$, $1.9\%$, $2.0\%$, $2.1\%$, and $2.2\%$, with
$\alpha=80\beta_e$. The $2.0\%$ point provides the branch anchor and is excluded
from the reported average; the other four points are recomputed with updated
orbit matrices, field blocks, shifted factorization, and eigensolution. The
parallel resolution is varied over $N_\theta=33$--129 on the periodic domain
$[-5\pi,5\pi]$. Both the two-field and three-field workloads use kinetic ions
and electrons with $E_{\max}/T_s=12.5$, $N_E=16$,
$N_{\mathrm{pitch}}=24$, and $N_b=24$.

At $N_\theta=97$, the mean changed-point times for the two-field workload are
$2.17$~s on the double-precision CPU, $0.127$~s on the double-precision GPU,
and $0.0574$~s on the single-precision GPU
[Fig.~\ref{fig:shen_electromagnetic_scan_performance}(a)]. These correspond to
reductions by factors of $17.0$ and $37.7$ for the two GPU paths relative to
the CPU path. For the three-field workload, the corresponding times are
$3.38$, $0.228$, and $0.0713$~s per changed point, giving factors of $14.9$ and
$47.5$ [Fig.~\ref{fig:shen_electromagnetic_scan_performance}(b)].

Adding $\delta B_\parallel$ enlarges the field Schur system and introduces its
associated field assembly and coupling blocks. At $N_\theta=97$, this raises
the measured cost relative to the two-field calculation by factors of about
$1.56$, $1.79$, and $1.24$ on the CPU double-, GPU double-, and GPU
single-precision paths, respectively. Nevertheless, both electromagnetic
field models retain subsecond changed-point throughput on the single-precision
GPU over the displayed resolution range.

\newpage
\bibliographystyle{unsrtnat}
\bibliography{references}

\end{document}